\documentclass[graybox, nosecnum]{svmult}

\usepackage{mathptmx}       % selects Times Roman as basic font
\usepackage{helvet}         % selects Helvetica as sans-serif font
\usepackage{courier}        % selects Courier as typewriter font
\usepackage{type1cm}        % activate if the above 3 fonts are
\usepackage{amssymb}        % not available on your system
\usepackage{makeidx}         % allows index generation
\usepackage{multicol}        % used for the two-column index
\usepackage[bottom]{footmisc}% places footnotes at page bottom
\usepackage{hyperref}        %for hyperlinks
\usepackage{soul}            % for high-lighting of text
\hypersetup{colorlinks=true,urlcolor=blue}

\usepackage{graphicx,array,multirow}	
\usepackage{amsmath}	
\usepackage{amssymb}	
\usepackage{bm}		
\usepackage[square,numbers]{natbib}
\makeindex             % used for the subject index
\begin{document}
%\tableofcontents{}
\title*{Massive Black Hole Mergers}
\author{Enrico Barausse \thanks{corresponding author} \& Andrea Lapi}
\institute{Enrico Barausse \at SISSA - Scuola Internazionale Superiore di Studi Avanzati, Via Bonomea 265, 34135 Trieste, Italy
\at IFPU - Institute for Fundamental Physics of the Universe, Via Beirut 2, 34014 Trieste, Italy
\at INFN-Sezione di Trieste, via Valerio 2, 34127 Trieste,  Italy
\at\email{barausse@sissa.it}
\and Andrea Lapi \at SISSA - Scuola Internazionale Superiore di Studi Avanzati, Via Bonomea 265, 34135 Trieste, Italy 
\at IFPU - Institute for Fundamental Physics of the Universe, Via Beirut 2, 34014 Trieste, Italy
\at INFN-Sezione di Trieste, via Valerio 2, 34127 Trieste,  Italy
\at INAF-Osservatorio Astronomico di Trieste, via Tiepolo 11, 34131 Trieste, Italy
\at \email{lapi@sissa.it}}
\maketitle
\abstract{At low redshift, massive black holes are found in the centers of almost all large elliptical galaxies, and also in many lower-mass systems. Their evolution is believed to be inextricably entangled with that of their host galaxies. On the one hand, the galactic environment provides gas for the black holes to grow via accretion and shine as active galactic nuclei. On the other hand, massive black holes are expected to backreact on the galactic dynamics, by injecting energy in their surroundings via jets or radiative feedback. Moreover, if galaxies and dark-matter halos form hierarchically, from small systems at high redshift coalescing into larger ones at more recent epochs, massive black holes may also merge, potentially generating gravitational-wave signals detectable by present and future experiments. In this Chapter, we discuss the predictions of  current astrophysical models for the mergers of massive black holes in the mHz frequency band of the Laser Interferometer Space Antenna (LISA) and in the nHz frequency band of pulsar-timing array experiments. We focus in particular on the astrophysical uncertainties affecting these predictions, including the poorly known
dynamical evolution of massive black hole pairs at separations of hundreds of parsecs; the possible formation of ``stalled'' binaries at parsec separations (``final-parsec problem''); and the effect of baryonic physics (e.g. SN feedback) on the growth of massive black holes. We show that nHz-band predictions
are much more robust than in the mHz band, and comment on the implications of this fact for LISA and pulsar-timing arrays.}

\section{Keywords} 
black holes --- gravitational waves --- dark matter halos --- galaxies: formation and evolution --- LISA --- pulsar-timing arrays

\section{Introduction}
In the local universe, massive black holes (MBHs) with  masses of $10^5$--$10^9 M_\odot$ are ubiquitous
in the center of large elliptical galaxies~\cite{Gehren84, Kormendy1995}, and are also present in some low-mass dwarf galaxies~\cite{reines11, reines13, Baldassare2019}. MBHs are also believed to accrete
from the nuclear gas and thus power Active Galactic Nuclei (AGNs) and quasars~\cite{2002apa..book.....F}, which are in turn expected to exert a feedback (via jets or radiation) on their surroundings and even on their  galactic host as a whole~\cite{Croton2006,2008ApJS..175..390H,2006MNRAS.370..645B}. As a result, the evolution of MBHs and galaxies is expected to proceed ``synergically'', as suggested also by the observed galaxy-MBH scaling relations~\cite{Kormendy2013, Mcconnell2013, Schramm2013,2013ApJ...768...76S,sahu1,sahu2}, which link e.g. the MBH mass to the stellar velocity dispersion of the host's spheroid ($M_{\rm bh}-\sigma$ relation) and the  MBH mass to the spheroid's stellar mass ($M_{\rm bh}-M^*$ relation). (See however \cite{VolonteriReines16, Shankar2016, Barausse2017, greene19} for recent work on these scaling relations, which shows that their interpretation may be  subtler and at least partially related to observational bias.)

Despite their crucial importance for galaxy formation and evolution, the mechanisms that drive the growth, feedback, and dynamics of MBHs are surprisingly little understood. From a theoretical prospect, this is to be ascribed to the difficulty of resolving the scale of MBHs and their sphere of influence, which is tiny compared to galactic scales. Moreover, processes such as star formation and supernova (SN) explosions, but also MBH accretion, feedback and generally the interaction of MBHs with the surrounding gas are not yet fully understood from first principles and have therefore to be modeled via ``sub-grid'' prescriptions in hydrodynamic simulations (cf. e.g. \cite{DiMatteo2005,2012MNRAS.420.2662D,Vogelsberger2014, Schaye2015, Volonteri2016,Tremmel2017, Pontzen2017, Nelson2019, Ricarte2019}). 

From an observational point of view, a crucial limitation consists in the difficulty of observing MBHs at moderate to high redshift, where their AGN/quasar activity is expected to be the brightest and their evolution the most rapid. Compared to electromagnetic probes, gravitational waves (GWs) offer several advantages in this respect (cf. Chapter 1). GWs interact very weakly with matter, unlike electromagnetic radiation. Moreover,  GW detectors are sensitive not to the energy flux (which decays as the inverse square of the luminosity distance $d_L$) but to the field (which decays as $d_L^{-1}$). For these reasons, GWs are in principle observable up to very high redshift.
GWs, unlike electromagnetic radiation, are also not produced microscopically, i.e. they are \textit{not} the collection of a huge number of quanta produced by atoms/plasmas. Instead, they are  produced macroscropically by the bulk motion of large masses moving at relativistic speeds. As such, they encode directly information on the dynamics and macroscopic properties of the system that generates them. For GWs resulting from the merger of two black holes, one can e.g. extract the masses and spins of the two objects from the GW signal alone (cf. Chapter 43).

Dark-matter (DM) halos and galaxies are believed to form hierarchically in the $\Lambda$ Cold Dark Matter ($\Lambda$CDM) model, from small systems at high redshift, which evolve to larger ones at lower redshift through a combination of major and minor mergers, as well as accretion of DM/hot gas from the intergalactic medium (IGM). It was therefore recognized early on~\cite{Begelman1980} that galaxy mergers are a promising environment for the production of GWs, because they could lead to the coalescence of the MBHs present at the centers of the two merging galaxies. Because of the large MBH masses, these black-hole mergers are not observable with current interferometers such as LIGO and Virgo (cf. Chapter 1). Indeed, the merger frequency of a black-hole binary system of total mass $M$ scales roughly as $\sim 1/(G M/c^3)$, which lies in the band $[10^{-4}\, {\rm Hz},0.1\,
{\rm Hz}]$ of the Laser Interferometer Space Antenna (LISA; cf. Chapter 3)
for masses $10^4$--$10^7 M_\odot$, and in the band 
$[10^{-9}\, {\rm Hz},10^{-7}\,
{\rm Hz}]$ of pulsar-timing arrays (PTAs; cf. Chapter 4) for masses
of $10^8$--$10^{9} M_\odot$. .

In this Chapter, we will review the astrophysics of the co-eval evolution of MBHs and their galactic hosts. We will start from the 
large scales  (i.e. $\lesssim 100$ kpc) of DM halos and the diffuse, chemically pristine intergalactic medium,
whose properties and merger history we will derive in detail within the framework of the $\Lambda$CDM model. We will then focus on the intermediate  scales (tens of kpcs down to pc) relevant for 
the baryonic physics of galaxies (e.g. stellar and gaseous disks and spheroids), down to the sub-pc scales of nuclear objects (nuclear star clusters and MBHs). We will then proceed to discuss the formation and evolution of MBHs (isolated and in binaries), starting from the formation of MBH pairs, which may then give rise to
bound binaries and eventually merge. We will discuss the uncertainties of this ``pairing'' phase of the MBH evolution, and present  its implications for LISA and PTAs.

Throughout this Chapter, we adopt the standard flat $\Lambda$CDM cosmology \cite{Planck2018} with rounded parameter values: fractional matter density at present time $\Omega_M\approx 0.3$, dark-energy density $\Omega_\Lambda\approx 0.7$, baryon density $\Omega_{\rm b}\approx 0.05$, Hubble constant $H_0 = 100\,h$ km s$^{-1}$ Mpc$^{-1}$  with $h\approx 0.7$, and mass variance $\sigma_8\approx 0.8$ on a scale of $8\, h^{-1}$ Mpc. 

\section{Dark matter halos}

In the standard cosmological framework, the seeds of cosmic structures like quasars, galaxies, and galaxy systems are constituted by DM perturbations of the cosmic density field, originated by quantum effects during the early inflationary universe. The perturbations are amplified by gravitational instabilities and, as the local gravity prevails over the cosmic expansion, are enforced to collapse and virialize into bound ``halos''. In turn, these tend to grow hierarchically in mass and sequentially in time, with small clumps forming first and then stochastically merging together into larger and more massive objects. The halos provide the gravitational potential wells where baryonic matter can settle in virial equilibrium, and via several complex astrophysical processes (cooling, star formation, feedback, etc.) originate the luminous structures that populate the visible universe (see  textbooks such as \cite{Cimatti2020,Mo2010}).

In the following, we recap, with a modern and original viewpoint, the crucial issues concerning the formation and evolution of DM halos, which will provide the backbone for the description of MBH mergers in the rest of this Chapter.

\subsection{Basic quantities}

Provided a background cosmology and suitable initial conditions in terms of a power spectrum $P(k)$ of density fluctuations (e.g., \cite{Bardeen1986}), the statistical evolution of the halo populations in mass and redshift can be characterized at leading order via three basic quantities: 

$\bullet$ The mass variance $\sigma(M)$, which describes the statistics of the density perturbation field when filtered on different mass scales $M$; this is defined as
\begin{equation}\label{eq|variance}
\sigma^2(M) = \frac{1}{(2\pi)^3}\,\int{\rm d}^3k\, P(k)\,\tilde{W}_M^2(k)~,
\end{equation}
where $\tilde{W}_M^2(k)$ is the Fourier transform of a window function whose volume in real space encloses the mass $M$. For a scale-invariant power spectrum $P(k)\propto k^n$ with effective spectral index $n>-3$ (to ensure hierarchical clustering), $\sigma(M)\propto M^{-(n+3)/6}$ applies.

$\bullet$ The linear threshold for collapse $\delta_c(z)$, which is a measure of the typical amplitude required for a perturbation to collapse efficiently; in fact, the condition $\sigma(M)\sim \delta_c(z)$ yields the characteristic mass $M_c(z)$ that, on statistical grounds, is prone to collapse at redshift $z$. A useful approximated expression for the collapse threshold in a flat $\Lambda$CDM cosmology is given by $\delta_c(z)=\delta_{c0}\,D(0)/D(z)$, in terms of the normalization \cite{Eke1996}
\begin{equation}\label{eq|deltac}
\delta_{c0} \simeq \frac{3}{20}\, (12\pi)^{2/3}\,\left[1+0.0123\log_{10}\Omega_M(z)\right]\approx 1.68 ~,
\end{equation}
and of the growth factor \cite{Weinberg2008}
\begin{equation}\label{eq|growth}
D(z) \simeq \frac{5}{2}\,\frac{\Omega_{M}(z)}{1+z}\,\left[\frac{1}{70}+\frac{209}{140}{\Omega_M(z)-\frac{1}{140}\,\Omega_M^2(z)+\Omega_M^{4/7}(z)}\right]^{-1}
\end{equation}
where $\Omega_M(z)=\Omega_M\,(1+z)^3/[\Omega_\Lambda+\Omega_M\,(1+z)^3]$; this latter redshift dependence is close to $D(z)\propto (1+z)^{-1}$ at $z\gtrsim 1$, and then slows down toward the present when the dark energy component kicks in.

$\bullet$ The virial overdensity $\Delta_{\rm vir}$, which  approximately renders the typical density contrast of perturbations at virialization (hence in the nonlinear regime). In a flat $\Lambda$CDM cosmology it can be computed via the approximated formula \cite{Bryan1998}
\begin{equation}\label{eq|deltavir}
\Delta_{\rm vir}\approx 18\,\pi^2+82\,[\Omega_M(z)-1]-39\,[\Omega_M(z)-1]^2~,
\end{equation}
with typical values $\Delta_{\rm vir}\approx 100$ at the present time and increasing toward $\Delta_{\rm vir}\approx 180$ for $z\gtrsim 1$. Related to this, the virial mass $M_{\rm vir}$ of a DM halo is defined as the mass contained within a radius $R_{\rm vir}$ inside which the mean interior density is $\Delta_{\rm vir}$ times the critical density $\rho_{\rm crit}\equiv 3\,H_0^2/8\pi\,G\approx 2.8\times 10^{11}\, h^2\, M_\odot$ Mpc$^{-3}$.

\begin{figure}[!t]
\centering
\includegraphics[width=\textwidth]{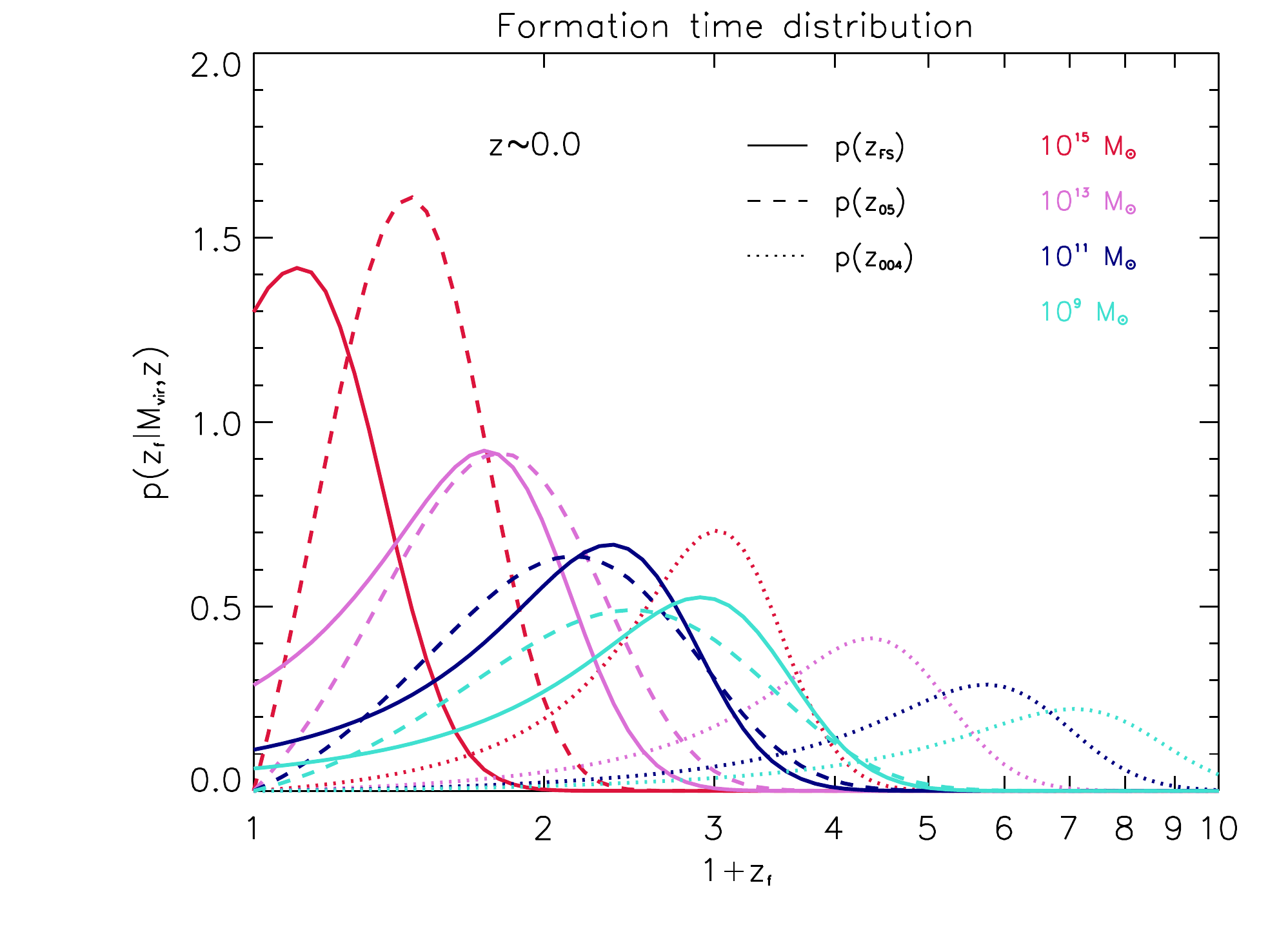}
\caption{Formation time distribution for different halo virial masses $M_{\rm vir}\approx 10^{15}\, M_\odot$ (red), $\approx 10^{13}\, M_\odot$ (orchid), $\approx 10^{11}\, M_\odot$ (blue), and $\approx 10^{9}\, M_\odot$ (cyan) at current redshift $z\approx 0$. Dotted lines refer to the redshift $z_{0.04}$ at which the halo has accumulated $4\%$ of its current mass, dashed lines to the redshift $z_{0.5}$ at which the halo has accumulated half of its current mass, and solid lines to the redshift $z_{\rm FS}$ of the transition between fast collapse and slow accretion.}\label{fig|probzform}
\end{figure}

\subsection{Formation time, fast collapse vs. slow accretion}

The formation redshift $z_f$ of a halo with mass $M$ at redshift $z$ is routinely defined as the highest redshift  at which the mass of its main progenitor is larger than $f M$, i.e. a fraction $f$ of the current mass. \cite{Giocoli2012} have provided a fitting formula of the formation redshift distribution extracted by $N-$body simulations, valid for any value of $f$. It writes
\begin{equation}\label{eq|zfdist}
\frac{{\rm d}p_f}{{\rm d}z_f}(z_f|M,z) = \frac{\alpha_f \, \nu_f\, e^{\nu_f^2/2}}{(e^{\nu_f^2/2}+\alpha_f-1)^2}\, \left|\frac{{\rm d}\nu_f}{{\rm d}z_f}\right|
\end{equation}
where $\alpha_f\approx 0.815\, e^{-2\, f^3}/f^{0.707}$ and
\begin{equation}\label{eq|wf}
\nu_f\equiv \frac{\delta_c(z_f)-\delta_c(z)}{\sqrt{\sigma^2(fM)-\sigma^2(M)}}~.
\end{equation}
The corresponding cumulative distribution is
\begin{equation}\label{eq|zfcumdist}
p_f(>z_f|M,z)= \frac{\alpha_f}{e^{\nu_f^2/2}+\alpha_f-1}~.
\end{equation}

Despite $f\approx 1/2$ is often exploited to define the formation redshift of a halo, this choice is rather arbitrary. In fact, intensive $N$-body simulations and analytic studies (see \cite{Zhao2003,Zhao2009,Lapi2011}) have demonstrated that the growth of a halo actually comprises two different regimes: an early fast collapse during which the central gravitational potential well is built up by dynamical relaxation processes; and a late slow accretion when mass is slowly added in the outskirts of the halo in the way of an inside out growth. The transition between the fast and slow accretion regime is a more motivated definition of halo formation; this is found to occur at a cosmic time
$t_{z_{\rm FS}}\approx 3.75\, t_{z_{0.04}}$, proportional to that at which the halo assembled a fraction $f\approx 4\%$ of its current mass. Thus, the generic formation time distribution ${\rm d}p/{\rm d}z_f$ in Eq. (\ref{eq|zfdist}) can be exploited to derive the distributions
${\rm d}p/{\rm d}z_{0.04}$ and ${\rm d}p/{\rm d}z_{\rm FS}$. Note that the latter can in principle extend even to redshifts $z_{\rm FS}<z$, meaning that a halo at redshift $z$ has not yet entered the slow accretion regime. The corresponding distributions for different halo virial masses $M_{\rm vir}\approx 10^9-10^{15}\, M_\odot$ at $z\approx 0$ are illustrated in Fig. \ref{fig|probzform}.

\subsection{Median and average halo mass growth}

The median history of the main progenitor for a halo with mass $M$ at redshift $z$ can be derived very easily from the formation redshift distribution Eq. (\ref{eq|zfdist}). 
By definition, if ${\rm d}p_f/{\rm d}z_f$ is the formation redshift distribution, and ${\rm d}p_{\rm MP}/{\rm d}M'$ is the distribution of main progenitor masses, one has the trivial identity
\begin{equation}\label{eq|basic}
\int_{z_f}^\infty{\rm d}z'\,\frac{{\rm d}p_{f}}{{\rm d}z'}(z'|M,z)=\int_{fM}^M{\rm d}M'\,\frac{{\rm d}p_{\rm MP}}{{\rm d}M'}(M',z_f|M,z)~.
\end{equation}
From this it is easily recognized that the cumulative distributions of formation redshift and main progenitor masses are equal; thus, both the median main progenitor mass at given redshift and the median formation redshift at given main progenitor mass both satisfy Eq. (\ref{eq|wf}), i.e.
\begin{equation}\label{eq|mahmed}
[\sigma^2(fM)-\sigma^2(M)]\, \tilde \nu_f^2 = [\delta_c(z')-\delta_c(z)]^2
\end{equation}
where
\begin{equation}
\tilde \nu_f = \sqrt{2\, \ln(1+\alpha_f)}
\end{equation}
is the median value that has been computed explicitly from Eq. (\ref{eq|zfcumdist}). The above equation can be solved for
the median $\tilde f(z')$ and hence the main progenitor median mass is $\tilde f(z') M$.

The average mass growth of the main progenitor $\langle f\rangle M$ can be related to the (cumulative) formation time distribution. By definition one has
\begin{equation}\label{eq|mahave}
\langle f(z')\rangle M = \int_{0}^M{\rm d}M'\,M'\,\frac{{\rm d}p_{\rm MP}}{{\rm d}M'}(M',z'|M,z)
\end{equation}
now one can rewrite $M'=\int_0^{M'}{\rm d}M''$, reverse the double (triangular) integral as $\int_{0}^M{\rm d}M'\,\int_0^{M'}{\rm d}M''=\int_{0}^M{\rm d}M''\, \int_{M''}^M{\rm d}M'$, and use Eq. (\ref{eq|basic}) to obtain
\begin{equation}
\langle f(z')\rangle M = M\, \int_0^1{\rm d}f\, p_f(>z'|M,z)
\end{equation}
Notice that to compute this quantity the formation time distribution for any $f$ is necessary.  

\begin{figure}[!t]
\centering
\includegraphics[width=\textwidth]{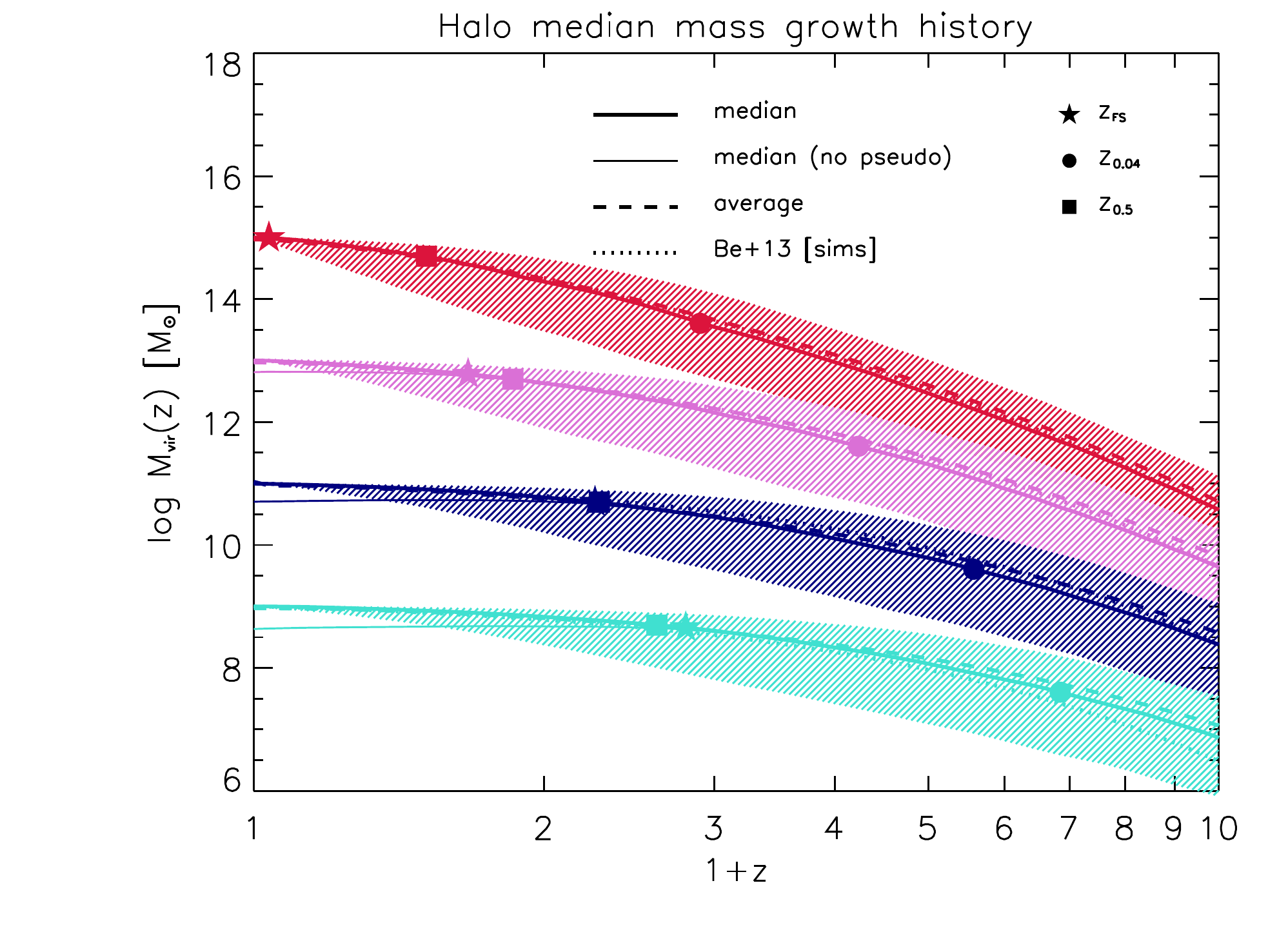}
\caption{Mass growth history of the main progenitor for halos with different virial masses $M_{\rm vir}\approx 10^{15}\, M_\odot$ (red), $\approx 10^{13}\, M_\odot$ (orchid), $\approx 10^{11}\, M_\odot$ (blue), and $\approx 10^{9}\, M_\odot$ (cyan) at current redshift $z\approx 0$. Thick solid lines refer to the median growth history, and shaded areas illustrate $5$th and $95$th percentile around it. Thin solid lines refer to the median history depurated from pseudo-evolution, dashed lines to the average growth history, and dotted lines show the direct simulation fits by \cite{Behroozi2013}. The symbols illustrate on every median growth history the location of the redshift $z_{0.04}$ (circles) where the main progenitor mass is $4\%$ of the current one, $z_{0.5}$ (squares) where it is $50\%$ of the current one, and $z_{\rm FS}$ (stars) where transition from fast to slow accretion regimes takes place.}\label{fig|mah}
\end{figure}

The corresponding median and average halo growth histories for different halo virial masses $M_{\rm vir}\approx 10^9-10^{15}\, M_\odot$ at $z\approx 0$ are illustrated in Fig. \ref{fig|mah} and found to be in overall good agreement with the tracks directly extracted from simulations \cite{Behroozi2013}. 

\subsection{Radial mass profile and pseudo evolution}

N-body simulations indicate that the mass profile of a DM halo follows the universal Navarro-Frenk-White (NFW) shape \cite{Navarro1997}
\begin{equation}\label{eq|NFW}
M(<r) = M_{\Delta}\, g(c_\Delta)\, \left[\ln\left(1+c_\Delta\, s \right)-\frac{c_\Delta\, s}{1+c_\Delta\, s}\right]
\end{equation}
where $s\equiv r/R_\Delta$ is the radius normalized to that $R_\Delta\equiv (M_\Delta/4\pi\, \Delta\rho_{\rm crit})^{1/3}$ within which the average density is $\Delta$ times the critical one $\rho_{\rm crit}\approx 2.8\times 10^{11}\, \Omega_M\,h^2\, M_\odot$ Mpc$^{-3}$; moreover, $c_\Delta$ is the concentration parameter and $g(x)\equiv [\ln(1+x)-x/(1+x)]^{-1}$. Very often $\Delta$ is taken to be the virial threshold for collapse $\Delta_{\rm vir}$ defined in Eq. (\ref{eq|deltavir}), so that the related concentration will be indicated by $c_{\rm vir}$.

\begin{figure}[!t]
\centering
\includegraphics[width=\textwidth]{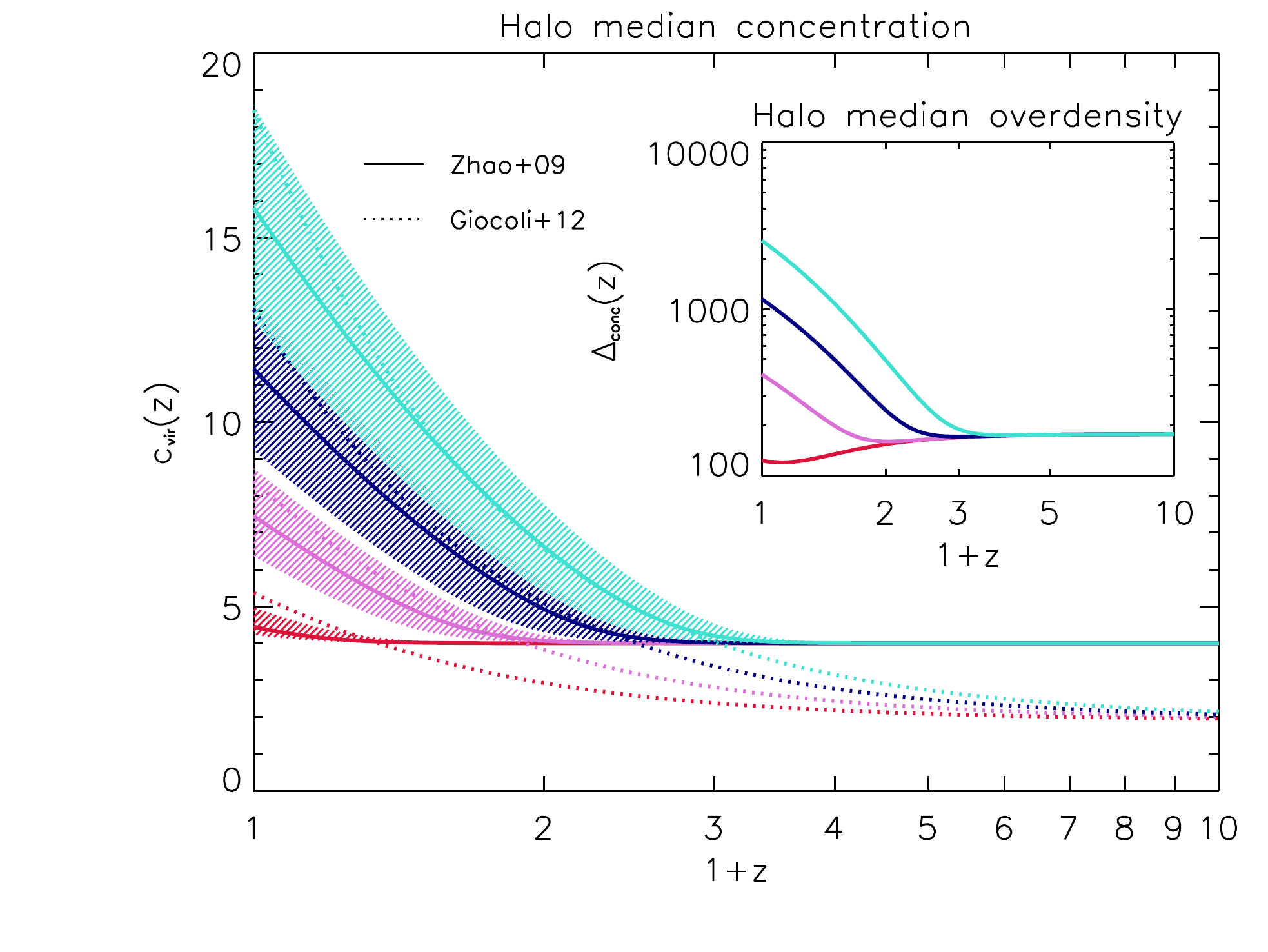}
\caption{Median virial concentration of halos with different halo virial masses $M_{\rm vir}\approx 10^{15}\, M_\odot$ (red), $\approx 10^{13}\, M_\odot$ (orchid), $\approx 10^{11}\, M_\odot$ (blue), and $\approx 10^{9}\, M_\odot$ (cyan) at current redshift $z\approx 0$; colored shaded area illustrate $5$th and $95$th percentile around the median. Solid lines refer to the prescription by \cite{Zhao2009} and dotted lines to that by \cite{Giocoli2012}. The inset shows the overdensity $\Delta_{\rm conc}$ of the inner region formed during the fast collapse phase.}\label{fig|conc}
\end{figure}

Simulations indicate (see \cite{Zhao2003,Zhao2009,More2015,Diemer2017}) that during the fast collapse the virial concentration $c_{\rm vir}$ stays put to a value around $4$, while it progressively increases as mass is added in the outer parts of the halo during the slow accretion phase. A simple fitting formula is given by \cite{Zhao2009}
\begin{equation}
c_{\rm vir}(z'|M_{\rm vir},z)\approx 4\, \left[1+(t_{z'}/t_{z_{\rm FS}})^{8.4}\right]^{1/8}
\end{equation}
in terms of the fast/slow transition time $t_{\rm FS}(M_{\rm vir},z)$; note that the latter depends on the halo virial mass $M_{\rm vir}$ at redshift $z$, and that actually is not a unique value but follows the distribution ${\rm d}p/{\rm d}z_{\rm FS}$ defined above. The median concentration corresponding to the halo mass growth histories discussed above for different halo virial masses $M_{\rm vir}\approx 10^9-10^{15}\, M_\odot$ at $z\approx 0$ is shown in Fig. \ref{fig|conc}.

It has been pointed out by several authors (e.g., \cite{Diemand2005,More2015,Diemer2017} that during the slow accretion phase the halo growth is mainly driven by a pseudo-evolution in radius and mass due to the lowering of the reference density defining the halo boundary. For example, pseudo-evolution implies that a halo of $10^{13}\, M_\odot$ with turning point at redshift $z_{\rm FS}\approx 2$ will end up, say, at $z\approx 0$ in a halo of several $10^{14}\, M_\odot$; clearly the latter is a halo typical of a galaxy group/cluster, but has little relevance to the physical processes occurring within the galaxy hosted by the original halo of $10^{13}\, M_\odot$ at formation. In other words, the `galactic' halo must have evolved in mass much less than predicted by pseudo-evolution. Under the assumption that the universal NFW mass distribution is retained, the mass growth depurated from pseudo-evolution can be estimated as \cite{More2015,Diemer2017}
\begin{equation}\label{eq|galhaloevo}
M_{\rm nops}(z)\simeq M_{\rm vir}(z)\, g[c_{\rm vir}(z)]/g[c_{\rm vir}(z_{\rm FS})]
\end{equation}
This also implies that the overdensity of the central region not subject to pseudo-evolution is given by
\begin{equation}
\Delta_{\rm nops}(z)\simeq \Delta_{\rm vir}\, \left[\frac{c_{\rm vir}(z)}{c_{\rm vir}(z_{\rm FS})}\right]^3\, \frac{g[c_{\rm vir}(z)]}{g[c_{\rm vir}(z_{\rm FS})]}
\end{equation}
We have highlighted in Figs. \ref{fig|mah} and \ref{fig|conc} the impact of the pseudo-evolution on the mass growth and on the concentration/central region overdensity.

\subsection{Halo spin}

DM halos slowly rotate with a specific (i.e., per unit mass) angular momentum given by 
\begin{equation}\label{eq|jhalo}
j_{\rm H}\approx 1670\,\frac{\lambda}{0.035}\,\left(\frac{M_{\rm H}}{10^{12}\, M_\odot}\right)^{2/3}\, E_{z}^{-1/6}~{\rm km~s^{-1}~kpc}~,
\end{equation}
where $E_z\equiv \Omega_\Lambda+\Omega_M\,(1+z)^3$ and $\lambda$ is the halo spin parameter. Numerical simulations (e.g., \cite{Bullock2001,Maccio2008,Zjupa2017}) have found that $\lambda$ has a roughly log-normal distribution with mean value $\langle\lambda\rangle\approx 0.035$ and dispersion $\sigma_{\log\lambda}\approx 0.25$ dex, approximately independent of mass and redshift. The small values of $\lambda$ testify that rotation is largely subdominant with respect to random motions, which indeed are mainly responsible for sustaining gravity and enforcing virial equilibrium.

The distribution of DM halo mass $M_{\rm H}(<j)$ as a function of the specific angular momentum $j$ has been studied in detail via $N$-body simulations by \cite{Bullock2001,vandenbosch2001,Sharma2005}, who found the convenient one-parameter representation
\begin{equation}\label{eq|jdist}
M_{\rm H}(<j) = M_{\rm H}\, \gamma[\alpha,\alpha j/j_{\rm H}]
\end{equation}
where $\gamma[a,x]\equiv \int_0^x{\rm d}t~t^{a-1}\, e^{-t}/\int_0^\infty{\rm d}t~t^{a-1}\, e^{-t}$ is the normalized incomplete gamma function, and the value $\alpha\approx 0.9$ applies. All in all, a representation in spherical mass shells $j\propto M(<r)^s$ with $s\sim 1.1-1.3$ holds to a very good approximation over an extended range.

\subsection{Halo mass function}

The halo mass function is the statistics describing the number of DM halos per unit comoving volume as a function of halo mass and redshift. This is routinely estimated via high-resolution, large-volume $N$-body simulations (see \cite{Sheth1999,Jenkins2001,Tinker2008,Bhattacharya2011,Watson2013}), although given the natural limits on resolution, computational time, and storing capacity, it can be probed only in limited mass and redshift ranges. We also caveat that the results of simulations depend somewhat on the algorithm used to identify collapsed halos (e.g., FoF=friend-of-friend vs. SO=spherical overdensity), and on specific parameters related to the identification of isolated objects (e.g., the linking length). 

\begin{figure}[!t]
\centering
\includegraphics[width=\textwidth]{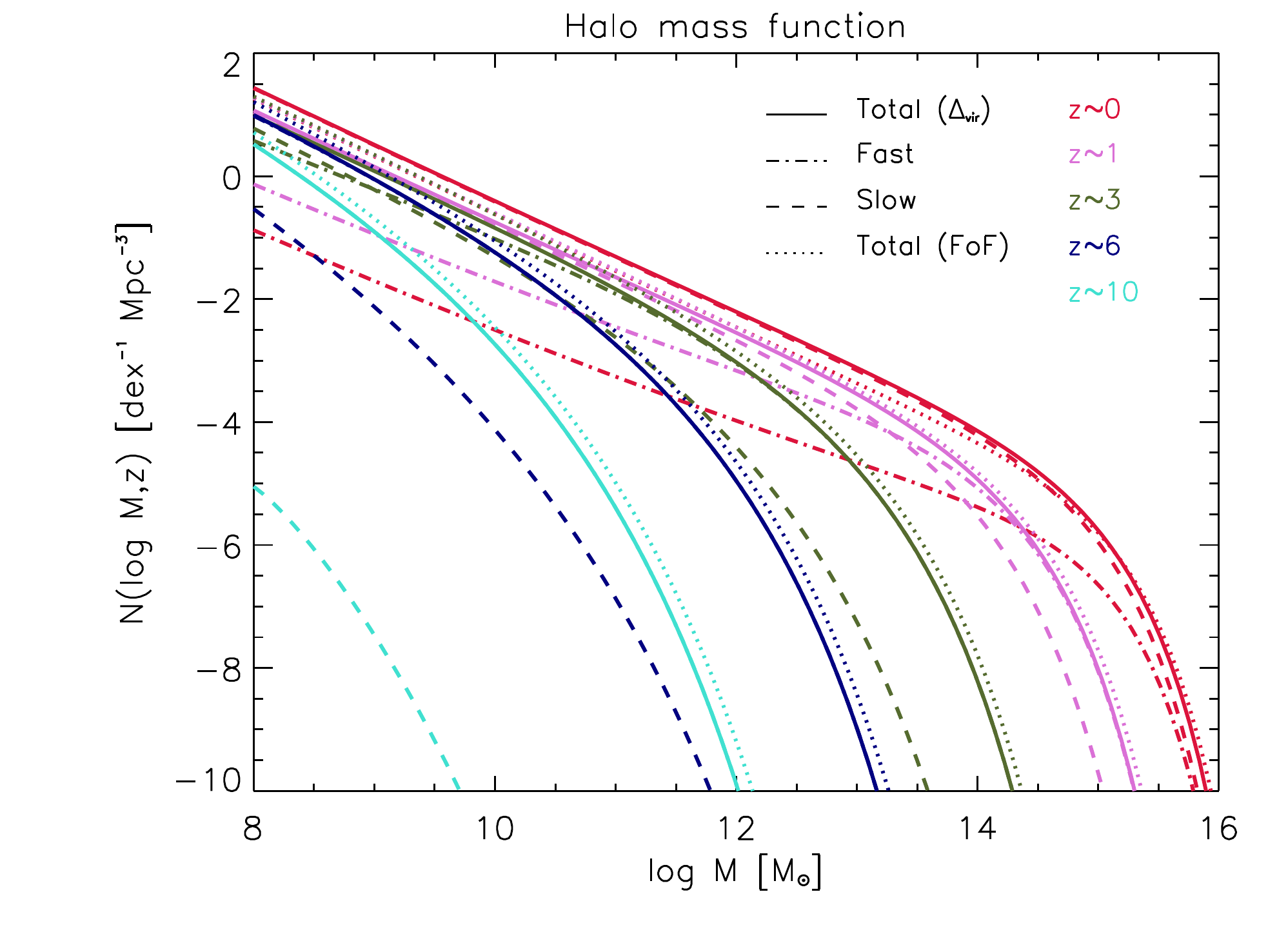}
\caption{Halo mass function $N(\log M,z)\equiv {\rm d}N/{\rm d}\log M$ per logarithmic mass bins (dex units) at different redshifts $z\approx 0$ (red), $1$ (orchid), $3$ (green), $6$ (blue) and $10$ (turquoise). Solid line is the total mass function by \cite{Watson2013} for a nonlinear threshold $\Delta_{\rm vir}$, dot-dashed line is the contribution from halos in the fast collapse stage, and dashed line from those in the slow accretion regime. Dotted lines illustrate the mass function for FoF-identified halos.}\label{fig|massfunc}
\end{figure}

One of the most accurate determination has been obtained by \cite{Watson2013}. Their fit to the simulation outcomes reads
\begin{equation}
\frac{{\rm d}N}{{\rm d}M} = \frac{\rho_{M}(0)}{M^2}\, \left|{\frac{{\rm d}\log \sigma}{{\rm d}\log M}}\right|\, \Gamma(\sigma,z|\Delta)\,f(\sigma,z|\Delta)
\end{equation}
where
\begin{equation}
f(\sigma,z|\Delta)=A\,\left[1+\left(\frac{b}{\sigma}\right)^a\right]\,e^{-c/\sigma^2}
\end{equation}
For FoF-identified halos the function $\Gamma(\sigma,z|\Delta)=1$ and the relevant parameters read $A=0.282$, $\alpha=2.163$, $\beta=1.406$ and $\gamma=1.210$; thus the function $f(\sigma)$ is actually independent of the redshift and the mass function is said to be universal. For halos identified with a spherical overdensity algorithm, one has
\begin{equation}
\Gamma(\sigma,z|\Delta)=C(\Delta)\, \left(\frac{\Delta}{178}\right)^{d(z)}\, e^{p\,(1-\Delta/178)/\sigma^q}
\end{equation}
and redshift-dependent parameters
$A=\Omega_M(z)\,[1.907\,(1+z)^{-3.216}+0.074]$, $a=\Omega_M(z)\,[3.136\, (1+z)^{-3.058}+2.349]$, $b=\Omega_M(z)\,[5.907\, (1+z)^{-3.599}+2.344]$, $c=1.318$,
$C(\Delta)=0.947\, e^{0.023\,(\Delta/178-1)}$, $d(z)=-0.456\, \Omega_M(z)-0.139$, $p=0.072$ and $q=2.130$. The halo mass function for both FoF and SO algorithms is illustrated in Fig. \ref{fig|massfunc}, together with the contribution from halo in the fast and slow accretion regime.

The shape of the mass function and other halo statistics (like progenitor mass function and large-scale bias) can be theoretically understood from first principles, on the basis of the celebrated Press \& Schechter theory by \cite{Press1974} and its extended version by \cite{Bond1991,Lacey1993,Mo1996}. Such a framework, modernly dubbed excursion set theory, remaps the issue of counting numbers of halos into finding the first crossing distribution of a random walk that hits a suitable barrier. The random walk is executed by the overdensity field  around a given spatial location when considered as a function of the mass variance $\sigma^2(M)$ of Eq. (\ref{eq|variance}). The barrier is provided by the linear collapse threshold $\delta_c(t)$ in Eqs. (\ref{eq|deltac}-\ref{eq|growth}), with possibly an additional dependence on the mass scale (‘moving barrier'). Remarkably, the resulting halo statistics are found to be in overall good agreement with the $N$-body outcomes, especially when a moving barrier with shape inspired by the ellipsoidal collapse is adopted \cite{Sheth1999,Zhang2006,Zhang2008}. Recently, an alternative framework to describe the halo statistics based on stochastic differential equations in real space has been proposed by \cite{Lapi2020}.

\subsection{Halo merger trees}

Halo merger trees are numerical yet approximate realizations of a halo merging history (see \cite{Kauffmann1993,Somerville1999,Cole2000,Parkinson2008}); these constitute the skeleton of many semi-analytic models aimed to describe the properties of galaxies and MBHs. 

\begin{figure}[!t]
\centering
\includegraphics[width=0.75\textwidth]{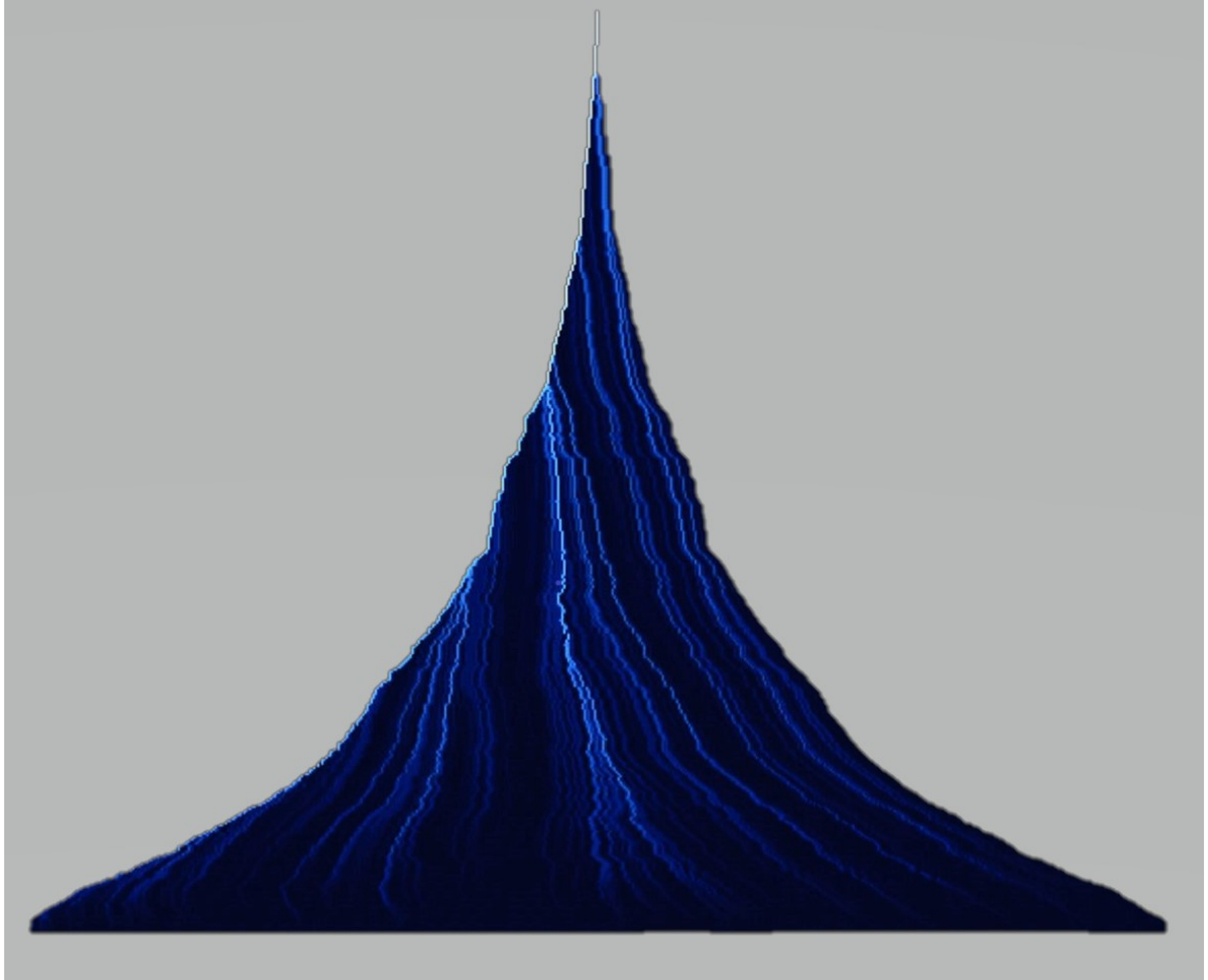}
\caption{Example of a merger tree for a descendant halo with final mass $10^{13}\, M_\odot$ at $z\approx 0$ (top), down to an initial redshift $z\sim 10$ (bottom); bluer colors correspond to more massive progenitors.}\label{fig|mergtree}
\end{figure}

The basic ingredient to build up the tree is the halo progenitor mass function ${\rm d}N/{\rm d}M'$, i.e., the distribution of halo masses $M'$ at different redshifts $z'$ that will end up in a given descendant halo mass $M$ at a later time $z$. This quantity is usually represented as
\begin{equation}
\frac{{\rm d}N}{{\rm d}M'}(M',z'|M,z) \simeq \frac{M}{M'}\,\frac{\Delta\delta_c\, e^{-(\Delta\delta_c)^2/2\Delta\sigma^2}}{\sqrt{2\pi}(\Delta\sigma^2)^{3/2}}\, G\left(1+\frac{\Delta\sigma^2}{\sigma^2},\frac{\delta_c^2}{\sigma^2}\right)\,\left|\frac{{\rm d}\sigma^2}{{\rm d}M}\right|_{M'}
\end{equation}
in terms of $\Delta\delta_c=\delta_c(z')-\delta_c(z)$ and $\Delta\sigma^2=\sigma^2(M')-\sigma^2(M)$; actually this expression
is inspired by the excursion set theory (see above), but includes a correction term to reproduce the results of N-body simulations \cite{Parkinson2008,Jiang2014}, that reads
\begin{equation}
G\left(1+\frac{\Delta\sigma^2}{\sigma^2},\frac{\delta_c^2}{\sigma^2}\right)
\simeq G_0\, \left(1+\frac{\Delta\sigma^2}{\sigma^2}\right)^{\gamma_1}\, \left(\frac{\delta_c^2}{\sigma^2}\right)^{\gamma_2}
\end{equation}
with $G_0\approx 0.57$, $\gamma_1\approx 0.19$ and $\gamma_2\approx -0.005$.

There are various possible algorithms to exploit the above and build up the tree, but one of the most effective in reproducing $N$-body outcomes is the \cite{Cole2000} binary method with accretion, which is briefly recalled next. First, provided a mass resolution $M_{\rm res}$, one computes the smooth mass accretion integrating the progenitor mass function
\begin{equation}
M_{\rm acc}\simeq \int_0^{M_{\rm res}}{\rm d}M'\, M'\, \frac{{\rm d}N}{{\rm d}M'}(M',z'|M,z)\,.
\end{equation}
Second, one computes the mean number of progenitors in the range $[M_{\rm res}, M/2]$ as
\begin{equation}
\mathcal{P}\simeq \int_{M_{\rm res}}^{M/2}{\rm d}M'\,\frac{{\rm d}N}{{\rm d}M'}(M',z'|M,z)\,,
\end{equation}
and chooses the merger time step so that $\mathcal{P}\ll 1$ to ensure that multiple fragmentation is unlikely. A uniform random number $\mathcal{R}$ generated in the interval $[0,1]$ determines whether the descendant has one ($\mathcal{R}>\mathcal{P}$) or two progenitors ($\mathcal{R}\leq\mathcal{P}$).
In the former case, one prescribes $M'=M-M_{\rm acc}$; in the latter case,
one of the masses $M'_1$ is drawn from the progenitor mass function in the range $[M_{\rm res}, M/2]$ and the second mass is just $M'_2=M-M'_1-M_{\rm res}$.

\begin{figure}[!t]
\centering
\includegraphics[width=\textwidth]{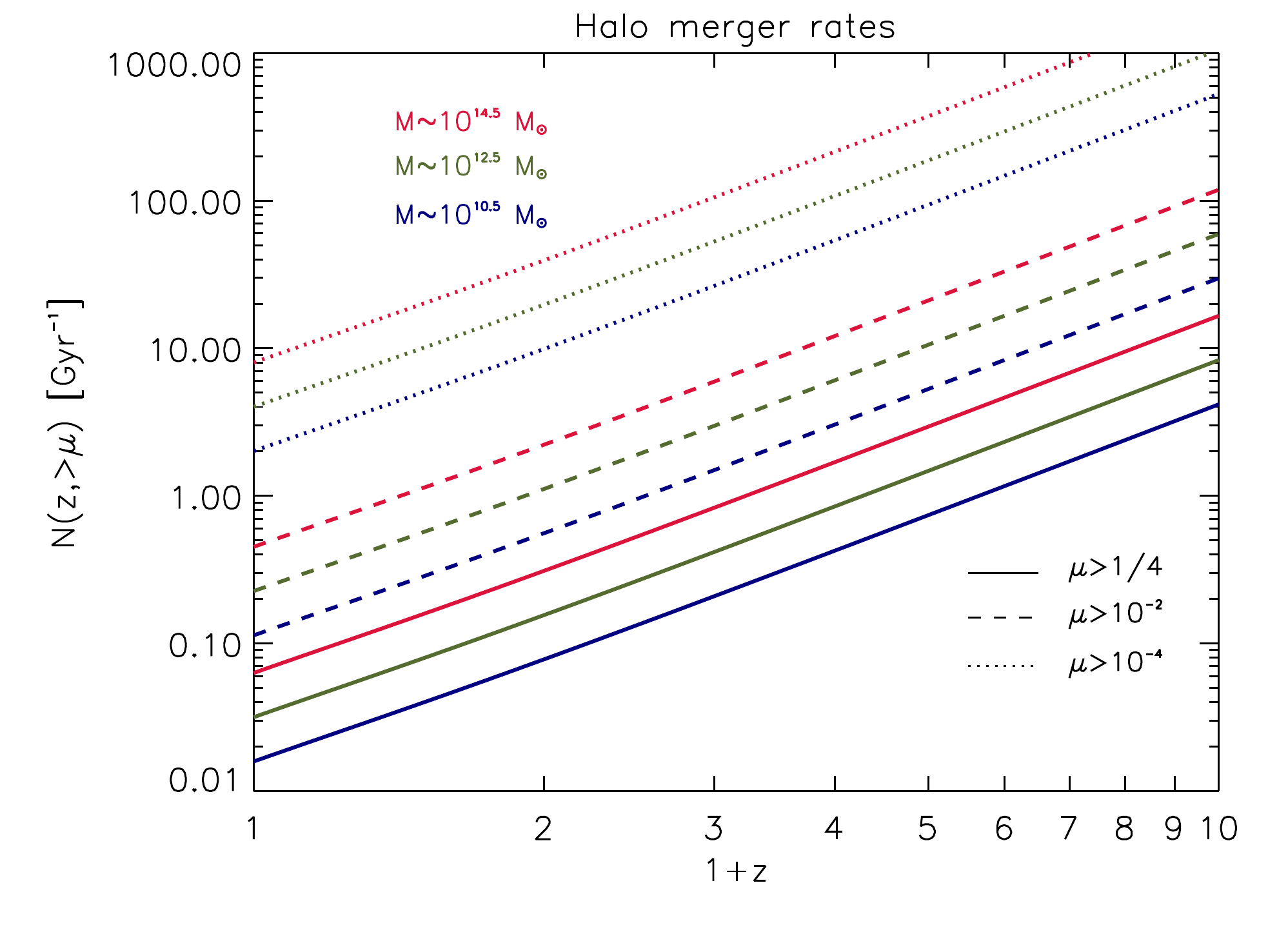}
\caption{Halo merger rate as a function of redshift for different halo masses $M_{\rm H}\approx 10^{14.5}\, M_\odot$ (red), $10^{12.5}\, M_\odot$ (green), $10^{10.5}\, M_\odot$ (blue), and different thresholds in halo mass ratios $\mu_{\rm H}>1/4$ (solid lines), $\mu_{\rm H}>10^{-2}$ (dashed lines), and $\mu_{\rm H}>10^{-4}$ (dotted lines).}\label{fig|halomergerates}
\end{figure}

\subsection{Halo merger rates}

After formation, a DM halo is expected to grow because of mergers and smooth accretion from the surrounding field and the cosmic web. The merger rates per descendant halo, per unit cosmic time $t$ (corresponding to redshift $z$) and per halo mass ratio $\mu_{\rm H}$ can be described with the fitting formula originally proposed by \cite{Fakhouri2010,Lapi2013}
\begin{equation}\label{eq|halomergerate}
\frac{{\rm d}N_{\rm H, merg}}{{\rm d}t\, {\rm d}\mu_{\rm H}} = N_{\rm H}\, M_{\rm H,12}^a\, \mu_{\rm H}^{-b-2}\, e^{(\mu_{\rm H}/\tilde\mu_{\rm H})^c}\, \frac{{\rm d}\delta_c}{{\rm d}t}
\end{equation}
in terms of the descendant halo mass $M_{\rm H,12}=M_{\rm H}/10^{12}\, M_\odot$, and of the linear threshold for collapse $\delta_c$. \cite{Genel2010} have determined the parameters entering the above expression from the \texttt{Illustris}-Dark simulations, finding $N_{\rm H}=0.065$, $a=0.15$, $b=-0.3$, $c=0.5$ and $\tilde\mu_{\rm H}=0.4$. Major mergers are typically identified with the events featuring $\mu_{\rm H}>1/4$, minor mergers with those featuring $1/10<\mu_{\rm H}<1/4$, and smooth accretion with those having $\mu_{\rm H}<1/100$.

The resulting halo merger rates as a function of the mass ratio for different redshifts are illustrated in Fig. \ref{fig|halomergerates}. We stress that these are often exploited to derive the MBH merger rates, by taking into account possible time delays due to a variety of dynamical processes; these will be discussed to some extent in the rest of this Chapter.

\section{Baryons and black holes}

\begin{figure}[!t]
\centering
\includegraphics[width=\textwidth]{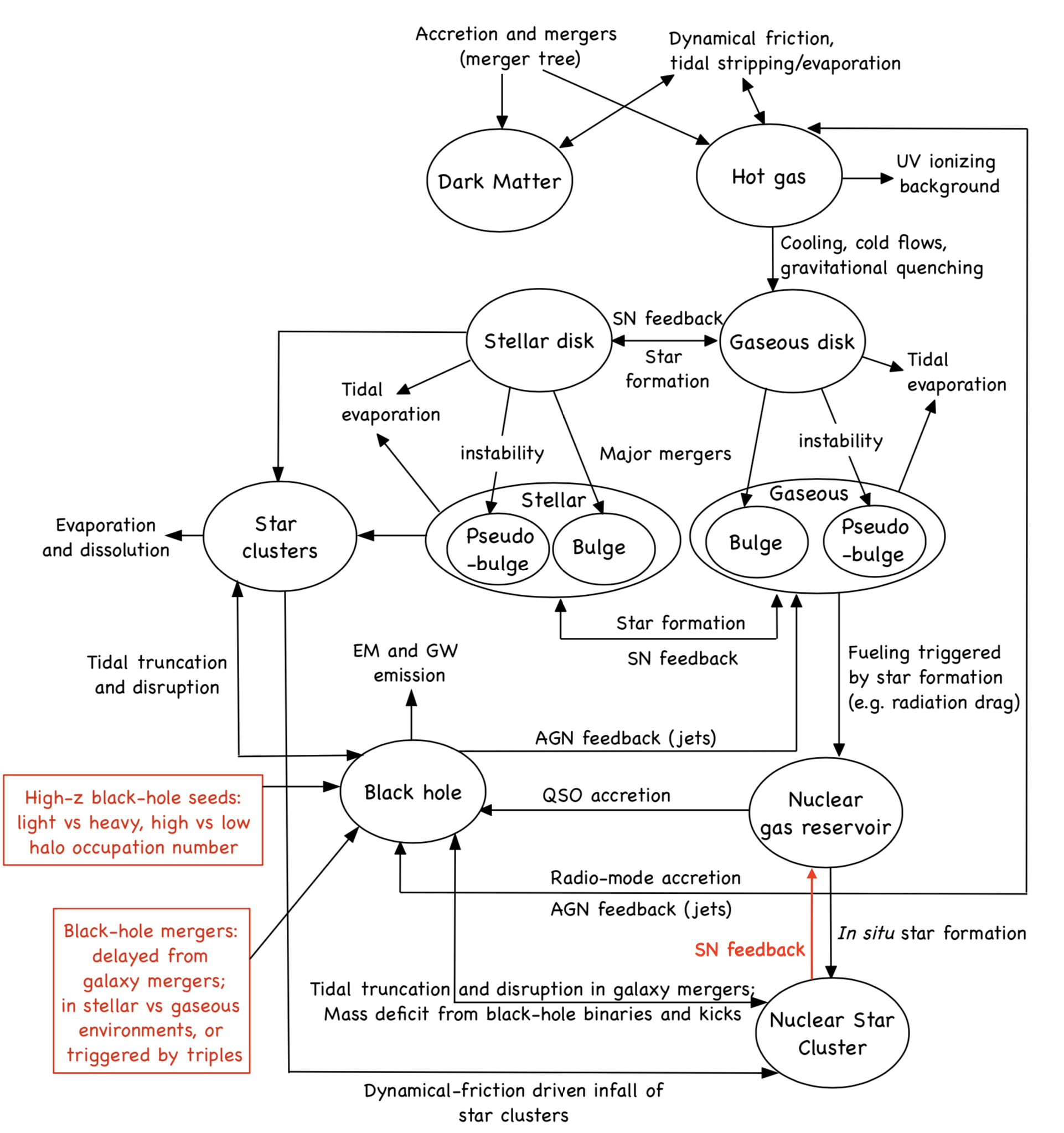}
\caption{Figure adapted from \cite{Klein2016}. Schematic description of the semi-analytic model of \cite{Barausse2012}, with the improvements of \cite{Sesana2014,Antonini2015,Bonetti2018b,2019MNRAS.486.4044B,EB2020}. Highlighted in red are the processes
 (black-hole seeding,  delays and SN feedback on the MBH growth) that most significantly affect MBH merger rates.\label{model}}
\end{figure}

The DM merger trees described above provide the backbone on top of which the hierarchical evolution of the baryonic structures of galaxies, as well as the MBHs, can be computed. In the following, we briefly describe the physics of baryons, before focusing on MBHs and presenting results for the MBH merger rates. We refer e.g. to \cite{Barausse2012} for a more extensive description of the physics. A schematic view of the various baryonic structures involved and their relations is presented in Fig.~\ref{model}.

On the largest scales,
baryons are mostly in the form of a chemically unprocessed gas that accretes onto DM halos from
the IGM. While accreting, the gas is
shock-heated to the virial temperature of the halo
in low-redshift systems of sufficiently large mass. However, in smaller systems and/or at high redshift, 
the IGM may simply stream into the DM halo  along cold flows/filaments~\citep{Dekel2006,Cattaneo2006,Dekel2009}. (In practice, in simulations, a smooth transition is observed  between these ``hot'' and ``cold'' accretion modes of the IGM, see e.g. \cite{correa}.)

Cooling of the shock-heated gas and/or  inflow of the IGM along cold filaments is then expected to give rise
to a cold-gas phase (``interstellar medium'', ISM). Because of conservation of angular momentum, this gaseous component may be disk-like~\cite{1998MNRAS.295..319M}, although its geometry can be significantly altered and rendered spheroidal by bar instabilities, compaction and (major) mergers. Moreover, the ISM (and particularly its molecular clouds) will also undergo star formation. The latter
will take place both in disks and in spheroids (or ``bulges''), although possibly more efficiently
in the latter~\cite{2010ApJ...714L.118D,2010MNRAS.407.2091G}, since the influx of gas toward the galaxy's center as a result of major mergers and disk instabilities may trigger starbursts.
Star formation will also exert a feedback on the stellar surroundings via SN explosions~\cite{fb1,fb2,fb3}, whose winds are expected to
expel gas from the ISM, thus quenching (or self-regulating) star formation. SN ejecta also  enrich the metal content of the ISM and contribute to its chemical evolution.

In their central regions, small to intermediate local galaxies with bulge velocity dispersion $\sigma \lesssim150$ km/s typically contain dense ($\sim 10$ pc) stellar clusters with mass
up to $10^8 M_\odot$~\cite{2006ApJ...644L..21F}. These ``nuclear star clusters'' 
might form from local (``in situ'') star formation episodes and/or infall of globular clusters from the galactic disk/bulge as a result of mass segregation (i.e. dynamical friction from the gas and field stars)~\cite{Antonini_Barausse2015,Antonini2015}. Their evolution is intimately connected with that of MBHs, which also dwell in the same nuclear region~\cite{graham1,graham2}. The merger of two galaxies, and the ensuing 
formation of a MBH pair or bound binary, is likely responsible for the relative scarcity of nuclear star clusters in systems with $\sigma \gtrsim150$ km/s~\cite{2010ApJ...714L.313B,Antonini_Barausse2015,Antonini2015}. Indeed, because of the galaxy-MBH scaling relations, those large galaxies tend to host large MBHs. When the latter form pairs/binaries, they  erode  nuclear star clusters by ejecting stars through the slingshot effect. Moreover,  MBH accretion occurs from the same nuclear gas (mainly resulting from inflows due to major mergers and/or disk instabilities) that feeds the growth of nuclear star clusters via in-situ star formation episodes. It is
therefore natural to expect SN explosions, which
tend to deplete/heat up the ISM,
to impact also the amount of gas that can accrete onto the MBH. This 
effect is indeed observed in hydrodynamic simulations, though it
 depends sensitively on the details of how the energy released by SNae couples to the ISM. For instance,
 weak thermal and kinetic SN feedbacks do not dramatically impact BH growth, but if the shocks caused by the explosions
 delay cooling of the gas, SN feedback may hamper BH
 growth in low-mass galaxies~\cite{Habouzit2017}.

Besides being passive actors in galaxy formation, with their mass growth and spin evolution proceeding through gas accretion and coalescences triggered by galaxy mergers, MBHs can also backreact on their galactic hosts. This process is known as AGN feedback~\cite{Croton2006,2008ApJS..175..390H,2006MNRAS.370..645B}. When accreting and shining as quasars/AGNs, MBHs emit photons in a variety of bands. A fraction of this radiation may interact with the (optically thick) material surrounding MBHs, heating it up and thus quenching its cooling and star formation. Moreover, AGN feedback may be provided by outflows/jets from the accretion disk (``disk winds'')~\cite{BP}, or by jets launched by the MBHs when they are highly spinning and immersed in an external magnetic field (anchored to a circum-nuclear disk). The latter process is known as Blandford-Znajek effect~\cite{1977MNRAS.179..433B}, is linked to the presence of a black-hole ergoregion, and is believed to be the mechanism underpinning radio-loud AGNs.

The baryonic physics that  we have described is typically implemented in both semi-analytic galaxy-formation models (e.g. \cite{sam1,sam2,Cole2000,sam4,Barausse2012}) as well as in hydrodynamic simulations (e.g. \cite{DiMatteo2005,2012MNRAS.420.2662D,Vogelsberger2014, Schaye2015, Volonteri2016,Tremmel2017, Pontzen2017, Habouzit2017, Nelson2019, Ricarte2019}). We will focus now on the two main uncertainties that affect the predictions for the merger rates of MBHs: the initial mass function of the MBH population at high redshift; and the time delays between the merger of two DM halos and the merger of the hosted MBHs.

\subsection{Black-hole mass function at high redshift}

Several formation mechanisms for the initial population of black-hole seeds at high redshift have been proposed (see e.g. \cite{Latif:2016qau} for a review). ``Light'' seeds (LSs) may form from the first generation (Population III) of stars~\cite{Madau2001}.
These metal-free stars are believed to form at high redshift (peaking around $z\sim 10$) in the deepest potential wells (i.e. in the most massive DM halos at a given epoch) and to be very massive (up to several hundred solar masses). 
When they explode as SNae, they may leave behind black  holes of a few hundred $M_\odot$, which may provide the seeds for the subsequent growth (via mergers and accretion) of the MBH population. 
A problem with this scenario is that it is difficult to reconcile it with the AGN luminosity function at high redshift, and particularly with the discovery of
active MBHs with masses $\gtrsim 10^9\, M_\odot$ 
in gas- and dust-rich  galaxies at $z\gtrsim 7$~\cite{2006AJ....131.1203F,2011Natur.474..616M,Banados_2018,2017ApJ...845..154V,2017ApJ...851L...8V,2018ApJ...866..159V}.
At  those redshifts the age of the universe was 
$\lesssim 0.77$ Gyr, and it seems difficult to accumulate those huge masses in such a short time span.

In more detail, if accretion onto these early LSs is   Eddington-limited, their mass grows
as $M_{\rm bh}\propto e^{t/\tau}$, with 
the  characteristic timescale $\tau$ given by
\begin{equation}
\tau = \frac{\eta}{(1-\eta)\, \lambda}\, t_{\rm Edd} \approx \frac{4.5\times 10^{7}}{\lambda}\, {\rm yr}\,.
\end{equation}
Here, $\lambda\equiv L/L_{\rm Edd}$ is the Eddington ratio between the black hole's bolometric luminosity  and the Eddington limit $L_{\rm Edd}\approx 1.4\times 10^{38}\, M_{\rm bh}/M_\odot$ erg s$^{-1}$; $t_{\rm Edd}=M_{\rm bh}\,c^2/L_{\rm Edd}\approx 0.4$ Gyr is the Eddington characteristic timescale; and $\eta\equiv L/(\dot M_{\rm bh}\, c^2)$ is the radiative efficiency
of the accretion flow, which we have set to $\eta=0.1$
(as approximately suitable for describing thin disks around slowly spinning black holes).
It is therefore clear that even for $\lambda\sim 1$ (Eddington accretion) a seed of a few $\times 10^2\, M_\odot$ would grow to $\gtrsim 10^9 M_\odot$ only at $z\sim 7$, which is marginally sufficient to explain the observations of \cite{2006AJ....131.1203F,2011Natur.474..616M,Banados_2018,2017ApJ...845..154V,2017ApJ...851L...8V,2018ApJ...866..159V}. Taking into account that 
accretion is also expected to proceed intermittently (in bouts possibly associated with major mergers and/or starbursts), it seems  that a LS formation scenario should require moderately super-Eddington accretion~\cite{Madau2014}. This is quite possible since at relatively high mass accretion rates $\dot{M}_{\rm bh}\gtrsim 0.5 L_{\rm Edd}/c^2$, accretion flows are expected to ``puff up'' and transition from a radiatively efficient, geometrically thin configuration to radiatively inefficient, geometrically slim disks, which can sustain moderately 
super-Eddington accretion~\cite{Madau2014}.

Another possibility is that MBH seeds may form already with relatively large masses $\sim 10^4$--$10^5M_\odot$ (see \cite{2019RPPh...82a6901M} for a recent review). Among these ``heavy seed'' (HS) scenarios are e.g. the rapid formation of seeds by direct collapse of gas and dust clouds in protogalaxies, induced by mergers, bar instabilities in gaseous disks, or cold-gas inflows along filaments~\cite{Volonteri2008,2010Natur.466.1082M,2015ApJ...810...51M,2012ApJ...745L..29D,2017MNRAS.467.4243D,2008MNRAS.387.1649B}; runaway 
collisions (favored by mass segregation) of massive
stars e.g. in metal-poor nuclear stellar
clusters~\cite{2012MNRAS.421.1465D,2016PASA...33...51L,2004Natur.428..724P}
or in the high-$z$, strongly star-forming galaxies with dense gas environments
that are progenitors to local early-type galaxies~\cite{Boco2020,2020MNRAS.498.5652K}.

\subsection{Delays between galaxy and black-hole mergers}

When two DM halos coalesce according to the merger-tree formalism
described earlier in this Chapter, the smaller one initially retains
its identity as a sub-halo (or ``satellite'') of the
newly formed system. That sub-halo then slowly spirals in (on typical timescales of a few Gyr) as  a result of dynamical friction~\cite{Boylan-Kolchin2008}. During this phase, tidal effects (stripping and evaporation) 
remove mass from both the DM and baryonic components, which 
in turn affects the evolution of the system (making dynamical friction less and less efficient)~\cite{Taffoni2003}. 

When the sub-halo finally reaches the center of the system, the baryonic components (the ``galaxies'')
and the contained MBHs 
do not coalesce immediately, but keep evolving under the same processes (dynamical friction and tidal stripping/evaporation)~\cite{EB2020,Dosopoulou2017,Tremmel2018}. This phase, during which the two MBHs go from $\sim$ kpc to $\sim$ pc separation, can also last for several Gyr, especially 
when the merging galaxies have  unequal stellar masses~\cite{Tremmel2018}.
Moreover, tidal effects  progressively
disrupt the smaller galaxy during its evolution, 
eventually leaving the MBH naked or at most surrounded by a core of stars~\cite{Dosopoulou2017}. 
As a result, a significant number of ``stalled'' MBHs  may be left wandering at
separations of hundreds of pc~\cite{Dosopoulou2017,Tremmel2018}.

For the MBHs that reach separations $\sim$ pc and form
bound binaries, dynamical friction eventually becomes inefficient compared to
other processes. Among the latter, a prominent role is played by stellar hardening~\cite{Quinlan1996,Sesana2015}, i.e. three-body interactions between the MBH binary and individual stars. Stars on low angular momentum orbits (i.e. in the ``loss cone'') interact  strongly with the binary, removing energy from it via the slingshot effect. Repeated interactions of this sort cause the MBH binary to shrink, while  stars in the loss cone are progressively removed from the system (being ejected e.g. as hypervelocity stars~\cite{Sesana2006}). As a result, the loss cone needs to be replenished (by diffusion of stars on the stellar relaxation timescale)  if stellar hardening is to efficiently
drive the binary's evolution down to separations $\sim 10^{-2}$--$10^{-3}$ pc (where GW emission alone can lead the system to coalescence in less than a Hubble time). Mechanisms that could help enhance stellar diffusion and thus replenish the loss cone (ensuring
hardening timescales of a few Gyr) include e.g.
triaxiality of the galaxy potential (resulting e.g. from a recent galaxy
merger)~\cite{Yu2002,2011ApJ...732...89K,Vasiliev2014,Vasiliev2014a,2015ApJ...810...49V} or galaxy rotation~\cite{Holley-Bockelmann2015}.

It should also be noted that in gas-rich galactic nuclei, planetary-like  migration
in a gaseous nuclear disk may harden the MBH binary on timescales even shorter ($\sim 10^7$--$10^8$ yr) than those of stellar interactions~\cite{MacFadyen2008,Cuadra2009,Lodato2009,Roedig2011,Nixon2011,Duffel2019,Munoz2019}. Moreover, even in gas-poor nuclei, and even if stellar hardening becomes inefficient due to insufficient loss-cone replenishment, 
 a third MBH will eventually be added to the system
by a later galaxy merger (as a consequence of the
hierarchical nature of structure formation,
described  in the first part of this Chapter)~\cite{Bonetti2018b,2019MNRAS.486.4044B,EB2020}.
Triple MBH interactions can trigger the merger of the inner binary via Kozai-Lidov oscillations~\cite{Kozai1962,Lidov1962}, which tend to decrease  the inclination of outer binary  while increasing the eccentricity of the inner one, or via chaotic
three-body interactions~\cite{Bonetti2016,Bonetti2018a}. Both processes can drive the inner binary  to
the GW-dominated regime (i.e. separations of  $\sim 10^{-2}$--$10^{-3}$ pc)
in a sizeable fraction of systems~\cite{Bonetti2018b,2019MNRAS.486.4044B,EB2020}.
Remarkably, this triple-MBH merger channel leaves a characteristic imprint on the GW signal observable by LISA, since MBH binaries originating from it are expected to carry a significant residual
eccentricity ($\gtrsim 0.99$ when they enter the LISA band, and $\sim 0.1$ at coalescence)~\cite{2019MNRAS.486.4044B}.

\begin{table}[b]
  \centering
  \caption{Predictions of the models of \cite{EB2020} for the total number of MBH mergers and detections  in $4$ years of observation with LISA. Adapted from \cite{EB2020}.}
  \label{tab:landscape}
  \begin{tabular}{|l||c|c|c|c||c|c|}
    \hline 
    \multirow{2}{*}{\textbf{Model}}  &  \multicolumn{2}{c|}{\textbf{LS}} & \multicolumn{2}{c||}{\textbf{HS}}   \\
    \cline{2-5}
                    & \textbf{Total} & \textbf{Detected} & \textbf{Total} & \textbf{Detected} \\
    \hline \hline
    \multicolumn{3}{c}{\emph{SN feedback}} \\ \hline
    SN-Delays & 48 & 16 & 25 & 25 \\\hline
    SN-shortDelays & 178 & 36 & 1269 & 1269  \\\hline\hline
    \multicolumn{3}{c}{\emph{No SN feedback}} \\ \hline
    noSN-Delays & 192 & 146 & 10 & 10 \\ \hline
    noSN-shortDelays & 1159 & 307 & 1288 & 1288   \\ \hline
    \hline
  \end{tabular}
  \label{tab:rates_new_names}
 \end{table}
 
 \section{Predictions for LISA and PTAs}
 
 \begin{figure}
\centering
\includegraphics[width=0.9\textwidth]{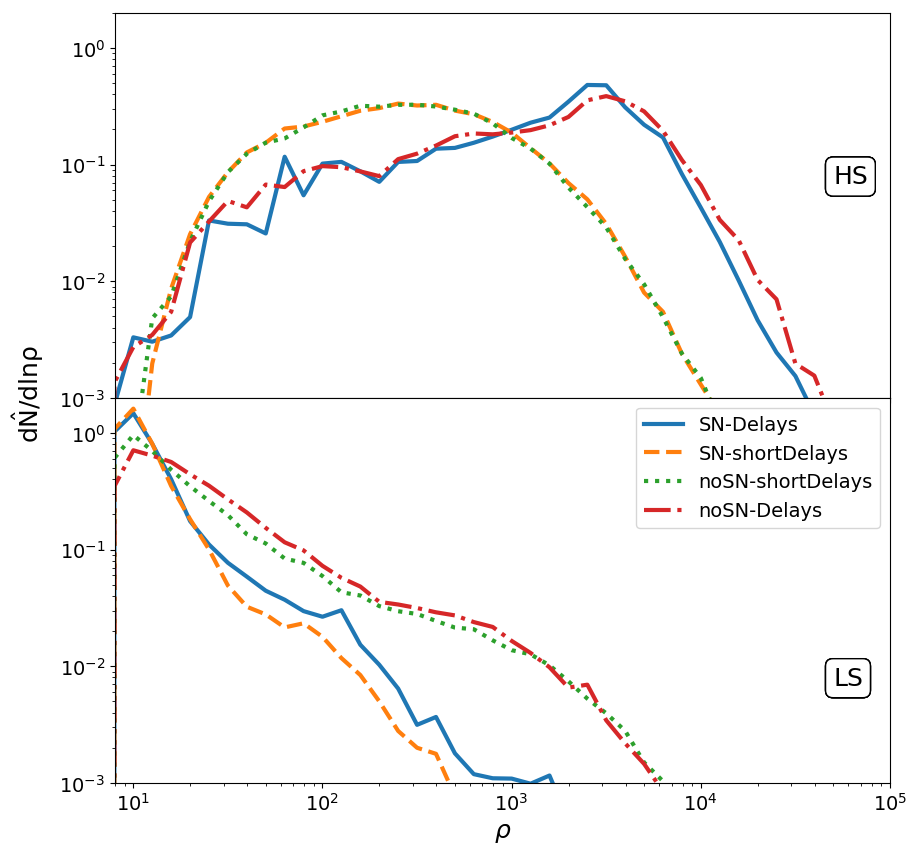}
\caption{Distribution of signal-to-noise ratio  (normalized to the total number of detections) for the HS and LS models (respectively upper and lower panels). Adapted from \cite{EB2020}.}
\label{fig:SN_dNdlSNR}
\end{figure}
 
 \begin{figure}
\centering
\includegraphics[width=0.9\textwidth]{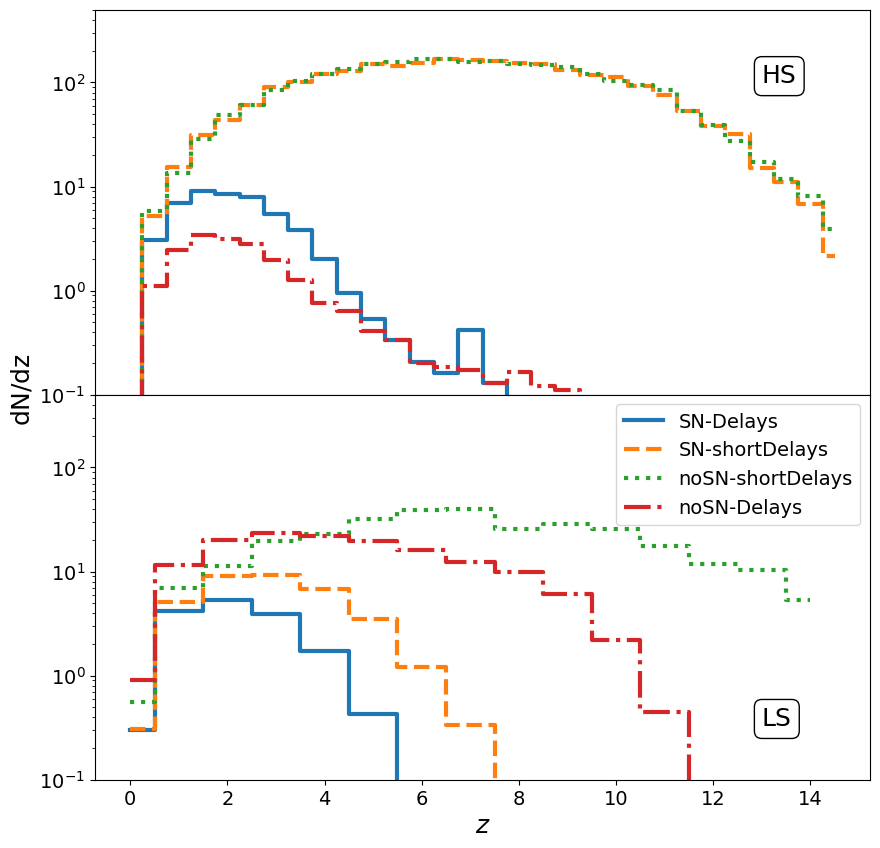}
\caption{Detected MBH mergers per unit redshift in 4 years of LISA observations,
for the HS models (upper panel) and the LS models (lower panel). Adapted from \cite{EB2020}.}
\label{fig:SN_dNdz}
\end{figure}

\begin{figure}
\centering
\includegraphics[width=0.9\textwidth]{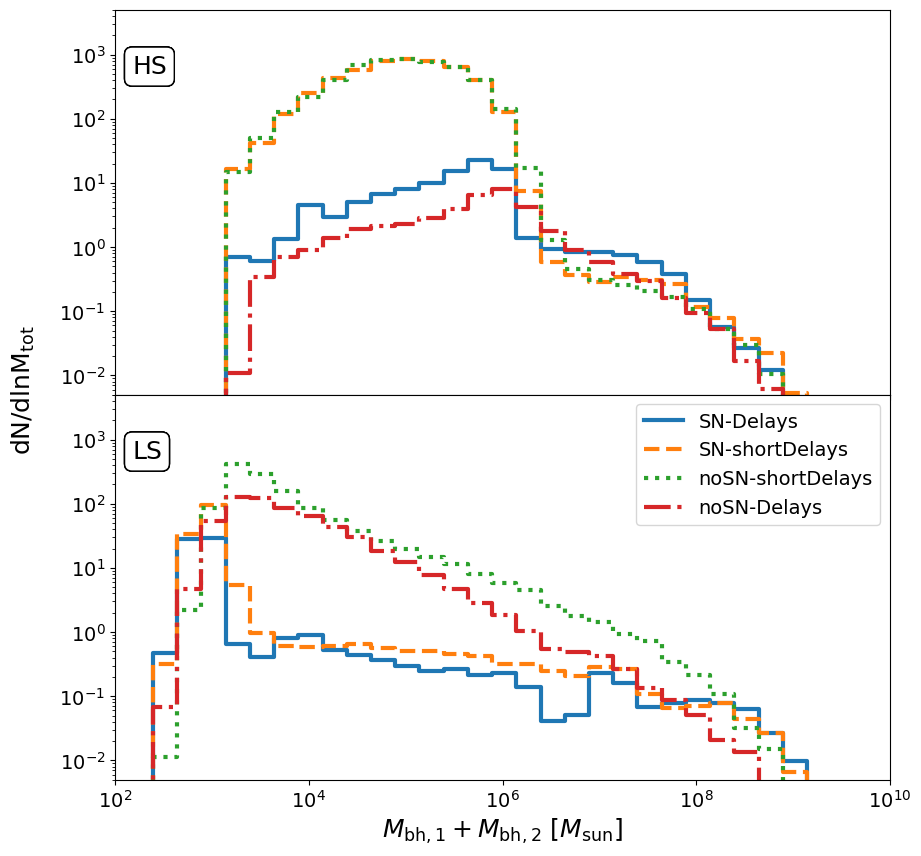}
\caption{Detected MBH mergers as a function of total MBH binary mass for the HS models (upper panel) and the LS models (lower panel).  Adapted from \cite{EB2020}.
}
\label{fig:SN_dNdM}
\end{figure}

\begin{figure}
\centering
\includegraphics[width=0.9\textwidth]{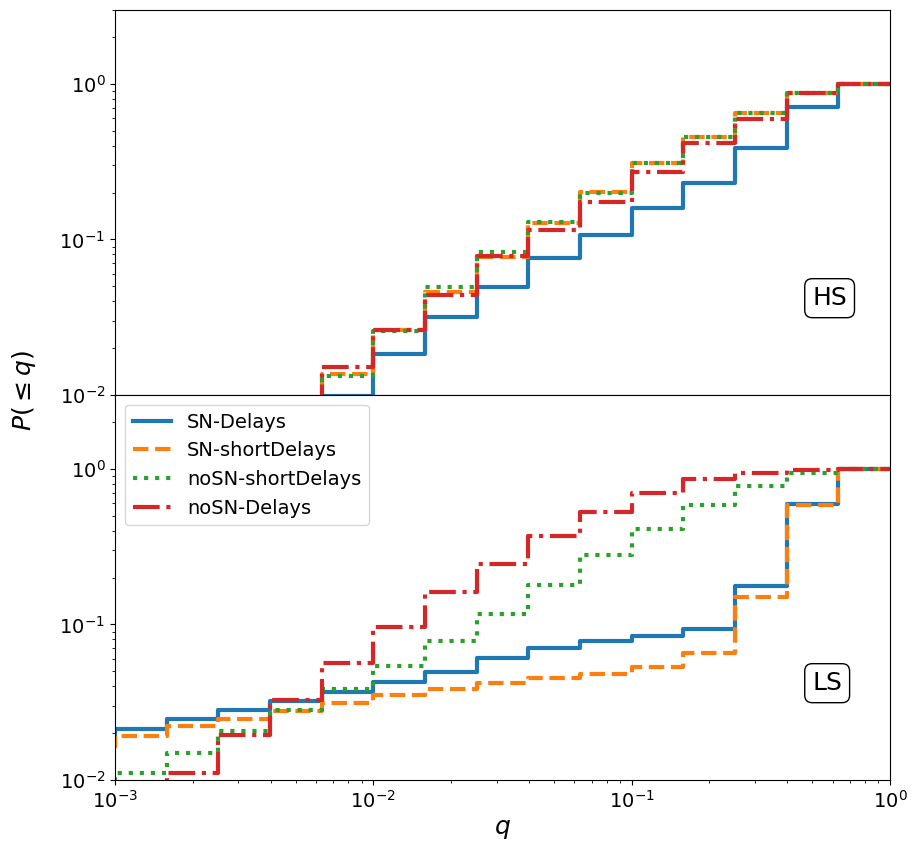}
\caption{Fraction of detected MBH mergers with mass ratio below $q$, for the HS models (upper panel) and the LS models (lower panel). Adapted from \cite{EB2020}.}
\label{fig:SN_models_dNdq}
\end{figure}

\begin{figure}
\centering
\includegraphics[width=0.9\textwidth]{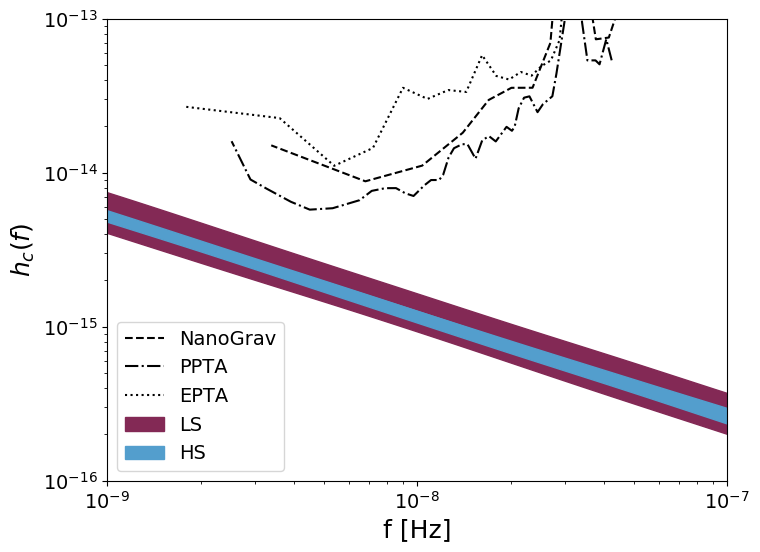}
\caption{Characteristic strain of the stochastic background from MBH binaries in the PTA band, for the eight models considered in this Chapter (with the envelope of the
LS/HS model predictions shown in purple/light blue, respectively).
Also shown are the sensitivities of ongoing   PTA experiments~\cite{Arzoumanian2018,Shannon2015,Desvignes2016}. Adapted from \cite{EB2020}.}
\label{fig:MBH_all_models}
\end{figure}
 
The predictions for how many MBH mergers will be observed by LISA, as well as for the parameters (masses, redshifts, etc) of those sources,
are very sensitive to the physics outlined above, and in particular to the sub-grid modeling of the seeding mechanism, the delays between galaxy and MBH mergers, and the impact of SN feedback on MBH growth.

To illustrate this fact, we will review here the results of the semi-analytic galaxy-formation model of \cite{EB2020} (based in turn
on the model of \cite{Barausse2012}, with the incremental improvements described in \cite{Sesana2014,Antonini2015,Bonetti2018b,2019MNRAS.486.4044B}  and in \cite{EB2020} itself). We will focus in particular on
two competing scenarios for the black-hole seeds, namely
 a population-III LS scenario~\cite{Madau2001},
and a representative HS model~\cite{Volonteri2008} where seeds
form (with masses
$\sim 10^4$--$\sim 10^5M_\odot$)
by direct collapse, as result of bar instabilities in high-redshift protogalaxies.
Moreover, besides considering versions of 
both the LS and HS models where
 SN feedback on the nuclear gas and  delays 
 are included (\textit{SN-Delays} models), we also review results obtained when either (or both) of these physical
 processes are switched off or modified. In particular, in the \textit{noSN} models we neglect the effect of SN explosions
 on the nuclear gas. In the \textit{shortDelays}
 models, we switch off the delays occurring as the MBH pair moves from
 kpc to pc separations (i.e. we neglect dynamical friction on the satellite galaxy and/or its MBH), while keeping the delays due to
 dynamical friction of the DM halos, as well as those due to stellar hardening, gas-driven migration, and triple MBH interactions.
 (Note that the  \textit{shortDelays} models correspond to the \textit{nodelays} models of \cite{EB2020}, since in that work the ``delays'' are only meant as those due to
 dynamical friction on the satellite galaxy and/or its MBH.)
 
 The (average)  number of MBH mergers detectable 
 by LISA in 4 years of observation is reported in Table~\ref{tab:rates_new_names}, alongside the
 total number of events (i.e. the number of mergers
 that could be detected in the same observation time
 if the detector had infinite sensitivity).
 We assume a signal-to-noise ratio ($\rho$) detection threshold $\rho>8$.
 As can be seen, 
 LISA will detect essentially all MBH coalescences in our past light cone
 in the HS models. That is the result of their large
masses, which produce larger signal-to-noise ratios (cf. also \cite{Klein2016}). Conversely, in the LS models, the fraction of detected MBH mergers 
is always less than one because of the lower binary masses, which translate into lower signal-to-noise ratios and GW frequencies at the high-end of the LISA sensitivity curve. The detection fraction is 
typically around 20--30\%, but can reach 
$\sim 75$\% in the \textit{noSN-Delays} (where  seed growth is unhampered and the delays
allow for longer MBH growth before binaries merge, yielding higher masses and signal-to-noise ratios).
Note also that the inclusion of realistic delays 
tends to decrease both the intrinsic number of MBH mergers and the number of detections. This  effect is more spectacular in the case of HSs.

The distribution of the signal-to-noise ratios of
the detected mergers is shown in Fig.~\ref{fig:SN_dNdlSNR}. As expected, 
mergers from HS models are typically louder 
than those from LS models, as a result of their 
larger masses. One can also see that
the differences between the HS models
are mainly due to the dynamics, with realistic delay
timescales resulting in louder sources (because 
longer delays allow the MBHs to grow to
larger masses, and shift mergers to lower redshifts, where
 signal-to-noise ratios are naturally higher).
In the LS models, the spread in the predictions is instead due mainly to SN feedback,
which lowers the typical mass of MBH binaries by
hampering their early growth.
Note that the relatively low signal-to-noise ratios
of the LS models highlight the importance of the LISA sensitivity level at high frequencies, where these
light sources are detected.

The distribution of  the detected mergers in redshift and in total mass of the MBH binary is shown in Figs.~\ref{fig:SN_dNdz}-\ref{fig:SN_dNdM}. Again, the effect of the delays,  which decrease the number of observed systems and shift the event distribution to lower redshift, is particularly evident. As for SN  feedback, it mainly affects LS models. Since it tends to prevent the MBHs from growing when they are located in  shallow potential wells at high redshift,
the redshift distribution peaks later in the LS models with SN feedback. Those LS models also present a noticeable
suppression in the number of mergers with masses below $\sim 10^7 M_\odot$, which is responsible for the 
lower number of detections in Table \ref{tab:rates_new_names}. Also note that
the low-mass peak in the distribution of the HS binaries in the models 
with short delays occurs because the MBH seeds do not have time to grow significantly
before merging.

Particularly important for  modeling  GW signals 
is the mass ratio of MBH binaries, whose distribution is
shown in Fig.~\ref{fig:SN_models_dNdq}.
As can be seen, most systems feature mass ratios
(defined as $q=M_{\rm bh, 2}/M_{\rm bh, 1}\leq1$) above $0.01-0.1$, according
to the specific model. Interestingly, the predicted distribution is very robust in HS models, while in  LS ones SN feedback tends to favor comparable-mass systems. This can be understood by noting that LS models with no  feedback allow seeds to grow more quickly away from
their initial masses, producing a larger range of possible mass ratios. When the seed growth is impeded by SN feedback, however, a significant number of mergers take place between black holes with masses close to their initial values, which are similar as they are produced by the same physical mechanism (explosions of population-III stars).

Finally, let us comment that while the number of  sources detected by LISA, as well as their parameters, is to a large extent model dependent, the predictions for the level of the stochastic background in the frequency band of PTAs are instead quite robust to the assumptions made about the seeding mechanism, the delays between galaxy and MBH mergers, and the effect of SN feedback on nuclear gas. This is evident from Fig.~\ref{fig:MBH_all_models}, which shows the predictions for the characteristic
strain $h_c$ of the stochastic background from MBH mergers in the $\sim\,$nHz frequency band,
for the eight models presented above,  compared with the 
 upper  bounds from EPTA, PPTA and NANOGrav~\cite{Arzoumanian2018,Shannon2015,Desvignes2016}.
 Note that the slope of the predictions 
 ($h_c\propto f^{-2/3}$)
 simply follows from the quadrupole formula of 
 general relativity, while the normalization (which depends
 on the MBH population)
  only shows a minor scatter according to the model.
  This robust dependence on the astrophysical model is 
  expected (cf. for instance also \cite{Bonetti2018b}),
  because the PTA signal comes from
  comparable-mass MBH binaries with 
    masses $\gtrsim 10^8 M_\odot$ at $z\lesssim 2$. The predictions
  for such systems depend very weakly on the initial seed mass function (since memory of the latter is lost due to accretion) and on the modeling of the delays between galaxy and MBH mergers (which for these systems are typically short).
  One physical ingredient that may affect the normalization of the PTA signal, however, is the overall normalization of 
  the black hole-galaxy scaling relations~\cite{Sesana2016}.
   The semi-analytic models presented in this Chapter were   calibrated to the observed scaling relations, accounting for the selection bias affecting the latter~\cite{Barausse2017}. In particular, as shown by \cite{Shankar2016}, 
   the black-hole sphere
of influence must be resolved for the MBH mass to be reliably estimated, which may bias the inferred normalization of the black hole-galaxy scaling relations.

\section{Future prospects}

While this Chapter was being finalized, NANOGrav
 reported evidence for a  common-spectrum
low-frequency stochastic process that affects 
pulsar-timing residuals in their 12.5-year dataset~\cite{Arzoumanian:2020vkk}.
The origin of this common ``red noise''
may be a stochastic GW background from MBH mergers. NANOgrav has so far reported no convincing evidence of
 quadrupolar angular correlations among the residuals of their pulsars (i.e. the Hellings-Downs correlation that should be present for a GW background~\cite{Hellings1983}). 
 However, if one interprets this  common ``red noise''
 as tentatively due to GWs, that would correspond to $h_c= A (f {\rm yr})^{-2/3}$, with 
 $A= 1.92\times 10^{-15}$ (median) and $5\%-95\%$ quantiles $A=1.37-2.67\times 10^{-15}$. This would be \textit{above} the previously published upper bounds shown in Fig.~\ref{fig:MBH_all_models}, although NANOgrav attributes this discrepancy to the (optimistic) use of log-flat priors in earlier analyses (including theirs).
 While it remains to be ascertained whether NANOgrav has really observed the stochastic background  from MBH binaries, it is possible that the
 predictions of this Chapter for  LISA  may need to be
 recalibrated (perhaps reconsidering the effect of selection effects on the scaling relations) and thus possibly revised to \textit{higher} event rates.
 In any case, even the models shown in Fig.~\ref{fig:MBH_all_models} would be testable by PTAs with
  $\approx 15-20$ yr of residual collection (assuming about 50 ms pulsars monitored at 100 ns precision). A much  earlier detection would  be 
  possible   with
the Square Kilometer Array (SKA) telescope~\cite{Dvorkin2017}.

\section{Acknowledgments}

E.B. acknowledges financial support provided under the European Union's H2020 ERC Consolidator Grant ``GRavity from Astrophysical to Microscopic Scales'' grant agreement no. GRAMS-815673. E.B. thanks I. Dvorkin, M. Bonetti, M. Tremmel and M. Volonteri for numerous insightful conversations on the astrophysics of galaxies and black holes, and for agreeing to adapting the figures of \cite{EB2020}. A.L. acknowledges financial support
from the EU H2020-MSCAITN-2019 Project 860744 `BiD4BEST: Big Data applications for Black hole Evolution STudies' and from the PRIN MIUR 2017 prot. 20173ML3WW 002, `Opening the ALMA window on the cosmic evolution of gas, stars and massive black holes'. We thank J. Gonzalez for proof-reading this manuscript.

\bibliographystyle{spbasic}
\bibliography{biblio}

\end{document}